\documentclass[aps,physrev,twocolumn,preprintnumbers,longbibliography,10pt]{revtex4-2}
\usepackage[colorlinks=true, linkcolor=blue, citecolor=blue, urlcolor=blue]{hyperref}
\usepackage{bm,latexsym,amssymb,amsmath,mathrsfs,mathtools,physics}
\usepackage{quantikz}
\usepackage{graphicx}
\usepackage{color}

\newcommand{\ri}{\mathrm{i}}

\begin{document}

\title{Quantum anomaly for benchmarking quantum computing}

\author{Tomoya Hayata}

\affiliation{Department of Physics, Keio University School of Medicine, Kanagawa 223-8521, Japan}
\affiliation{Center for Interdisciplinary Theoretical and Mathematical Sciences (iTHEMS), RIKEN, Saitama 351-0198, Japan}
\affiliation{International Center for Elementary Particle Physics, The University of Tokyo, Tokyo 113-0033, Japan}

\author{Arata Yamamoto}

\affiliation{Center for Interdisciplinary Theoretical and Mathematical Sciences (iTHEMS), RIKEN, Saitama 351-0198, Japan}

\preprint{RIKEN-iTHEMS-Report-26}

\begin{abstract}
Given the rapid advances in quantum computing hardware, establishing strategies for verifying the correctness of quantum computations has become increasingly important.
Exploiting the fact that the axial anomaly in gauge theories is exact to all orders in perturbation theory, we propose the axial anomaly as a nontrivial benchmark for quantum simulations of lattice gauge theories.
We simulate axial charge production in ${\mathbb Z}_N$ lattice gauge theories on the trapped-ion quantum computer ``Reimei''. 
After taking the U(1), infinitesimal time, and infinite volume limits, we successfully reproduce the anomaly coefficient within statistical uncertainties, even without error mitigation.
Our results demonstrate that the axial anomaly can be simulated on current quantum computers and serves as a verification test of quantum computations.
\end{abstract}

\maketitle

\section{Introduction}

Quantum computing is expected to play a central role in the future development of quantum field theory. 
Quantum hardware has rapidly advanced and the number of qubits in available devices has increased year by year, although harnessing their full performance remains challenging due to device noise. 
For small-scale quantum computations, their validity can be established through comparison with exact solutions obtained on classical computers \cite{Klco:2019evd,Yamamoto:2020eqi,Nguyen:2021hyk,Mildenberger:2022jqr,ARahman:2022tkr,Atas:2022dqm,Pomarico:2023png,Charles:2023zbl,Angelides:2023noe,Cochran:2024rwe,Davoudi:2025rdv,Rosanowski:2025nck,Hayata:2026rmv}. 
On state-of-the-art quantum devices, however, the number of qubits has exceeded one hundred, rendering exact calculations infeasible for classical computers. 
In this regime, validation must rely on comparisons with approximate methods, such as tensor network approaches \cite{Farrell:2023fgd,Farrell:2024fit,Hayata:2024smx,Hayata:2024fnh}.
While previous quantum computations of lattice gauge theories are finite dimensional \cite{Klco:2019evd,Yamamoto:2020eqi,Nguyen:2021hyk,Mildenberger:2022jqr,ARahman:2022tkr,Atas:2022dqm,Pomarico:2023png,Charles:2023zbl,Angelides:2023noe,Cochran:2024rwe,Davoudi:2025rdv,Rosanowski:2025nck,Hayata:2026rmv,Farrell:2023fgd,Farrell:2024fit,Hayata:2024smx,Hayata:2024fnh}, appropriate limits must eventually be taken to study quantum field theory with an infinite-dimensional Hilbert space.
Thus, benchmarking quantum computations for quantum field theory is much more challenging.
An ideal test case would be nontrivial (i.e., interacting) but exactly solvable in the infinite-dimensional Hilbert space, accessible to current devices, and scalable to future larger devices. 
Quantum field theory provides such a benchmark: the axial anomaly.

The axial anomaly relation is expressed as $\partial_\mu j_5^\mu(x) = \frac{1}{\pi}eE(x)$ in quantum electrodynamics in $1+1$ dimensions \cite{Abdalla:1991vua}.
This operator relation is exact to all orders in perturbation theory.
In particular, the coefficient $\frac{1}{\pi}$ is one-loop exact and remains protected against higher-order corrections.
Consider externally applying a spatially uniform electric field $E_{\rm ext}$.
The classical contribution dominates the right-hand side and quantum corrections are of higher order and drop off. 
Taking the expectation value of the above relation and integrating over a periodic box, we get
\begin{equation}
\label{eqAC}
  \frac{d}{dt} \langle Q_5\rangle = \frac{1}{\pi} L eE_{\rm ext}
\end{equation}
with the box size $L$.
This relation provides a direct benchmark for quantum computation.
The left-hand side can be evaluated by simulating the time evolution of the axial charge $Q_5$.
By varying the magnitude of the external electric field and computing the left-hand side, we can extract the coefficient on the right-hand side.
If the extracted coefficient matches the correct value $\frac{1}{\pi}$, we can conclude the correctness of the quantum computation.
Quantum electrodynamics is infinite dimensional but each simulation is finite dimensional.
Equation \eqref{eqAC} is reproduced only if all the necessary limits are successfully taken.
In this sense, this benchmark is not simply a check for a single simulation, but a verification that infinite-dimensional quantum field theory can be quantitatively reproduced from finite-dimensional simulations on an actual device.

The axial anomaly in lattice gauge theory has a long history.
Originally, the Nielsen–Ninomiya theorem posed a fundamental obstacle to formulating axial symmetry on lattices \cite{Nielsen:1981hk}.
After the discovery of a Hamiltonian formulation with exact axial symmetry \cite{Creutz:2001wp}, theoretical understanding of the axial anomaly has advanced \cite{Matsui:2005uh,Hayata:2023zuk,Chatterjee:2024gje,Hidaka:2025ram,Onogi:2025xir,Aoki:2025vtp}.
In particular, it has been recognized that a conserved axial charge in $1+1$ dimensions can be locally defined and simplifies practical calculations of the axial anomaly \cite{Hayata:2023skf,Singh:2025sye}.
Because time evolution is obstructed by the sign problem in classical Monte Carlo simulations, a direct simulation of Eq.~\eqref{eqAC} constitutes not only a validity test of quantum computation but also a longstanding challenge in lattice gauge theory.

\section{Theoretical background}

The axial anomaly is expected in U(1) gauge theory, but U(1) gauge theory requires an infinite-dimensional Hilbert space.
We formulate ${\mathbb Z}_N$ lattice gauge theories with finite-dimensional Hilbert spaces, and then take the U(1) limit $N \to \infty$.
We work in lattice units, where the lattice spacing is set to one, for notational simplicity.

\begin{figure*}[t]
    \centering
    \includegraphics[width=\linewidth]{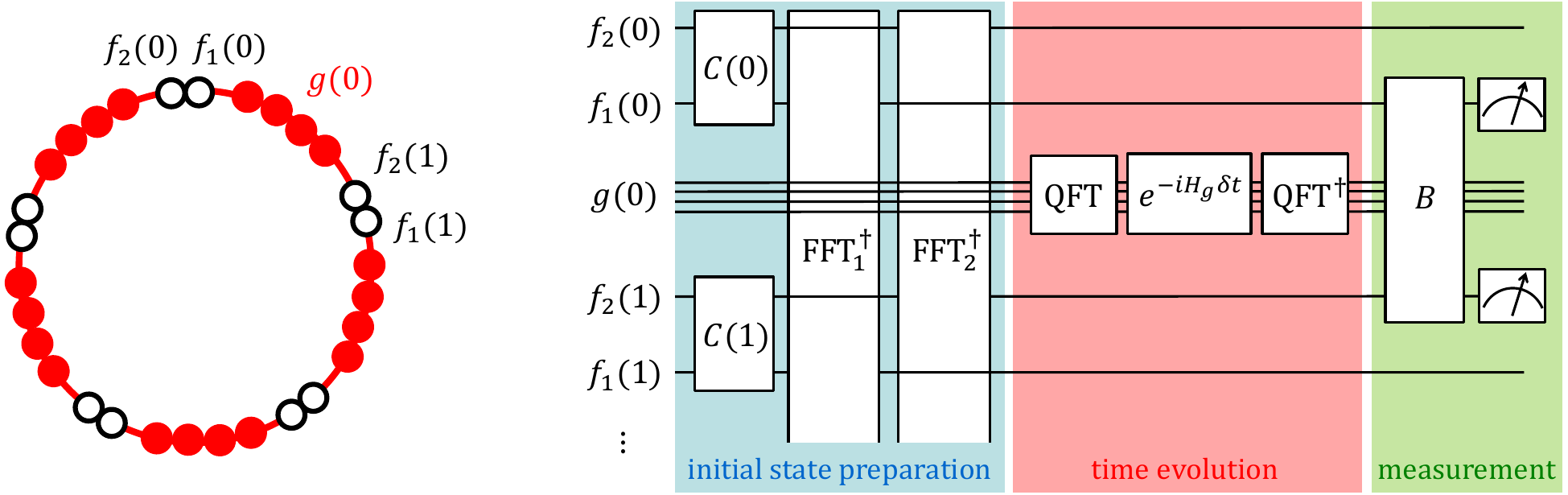}
    \caption{\label{figschematic}
    Left: lattice geometry.
    The two-component fermions $|f_1(x)\rangle$ and $|f_2(x)\rangle$ are defined on sites (black open circles).
    The gauge fields $|g(x)\rangle$ are defined on links (red filled circles).
    Periodic boundary conditions are imposed.
    Right: quantum circuit.
    The first stage is the initial state preparation implemented by the free fermion creation $C(k)$ and the inverse fermionic Fourier transform ${\rm FFT}_i^\dagger$.
    The second stage is the time evolution $e^{-\ri H_g \delta t}$ with the quantum Fourier transform QFT and its conjugate ${\rm QFT}^\dagger$.
    The third stage is a projective measurement after the basis transformation $B$ to the measurement basis.
    }
\end{figure*}

The Hamiltonian is given by the sum of the gauge and fermion parts, $H = H_g + H_f$.
In ${\mathbb Z}_N$ gauge theory, the $N$-dimensional gauge link operator $U(x)$ and the conjugate operator $\Pi(x)$ obey the commutation relation $\Pi(x)U(x)=e^{\ri \frac{2\pi}{N}} U(x)\Pi(x)$ \cite{Horn:1979fy}.
The gauge field Hamiltonian is
\begin{equation}
 H_g = -\frac{e^2}{2} \sum_{x} \left\{ \Pi(x) + \Pi^\dagger(x) \right\}
\end{equation}
with the gauge coupling constant $e$.
This becomes the standard electric field Hamiltonian, up to an irrelevant constant, in the U(1) limit.
The Wilson fermion is adopted for its lower computational cost compared with the Kogut-Susskind fermion.
The two-component fermion creation and annihilation operators satisfy $\{\psi_i^\dagger(x),\psi_j(x)\}=\delta_{ij}$.
With the gamma matrices $\gamma^0=\sigma_1$ and $\ri\gamma^1=\sigma_3$, the massless fermion Hamiltonian takes a simple form
\begin{equation}
\begin{split}
  H_f &= \sum_x \bigg\{ \psi^\dagger_1(x) \psi_2(x) + \psi^\dagger_2(x) \psi_1(x) \\
  &\quad - \psi^\dagger_1(x) U(x) \psi_2(x+1) - \psi^\dagger_2(x+1) U^\dagger(x) \psi_1(x) \bigg\}
\end{split}
\end{equation}
on a one-dimensional chain illustrated in the left panel of Fig.~\ref{figschematic}.
The theory is formulated on a one-dimensional lattice with periodic boundary conditions.

This Hamiltonian has two charge operators.
One is the vector charge
\begin{equation}
 Q = \sum_x \left\{ \psi^\dagger_1(x) \psi_1(x) + \psi^\dagger_2(x) \psi_2(x) -1 \right\}.
\end{equation}
The vector charge commutes with the total Hamiltonian, which leads to the conservation law
$\frac{d}{dt} Q = \ri [H,Q] = 0$.
The other is the axial charge 
\begin{equation}
\label{eqQ5}
    Q_5 = Q_M + Q_U,
\end{equation}
which is decomposed into the $U$-independent term
\begin{equation}
    Q_M = \sum_x \frac{\ri}{2} \left\{ - \psi^\dagger_1(x) \psi_2(x) + \psi^\dagger_2(x) \psi_1(x) \right\}
\end{equation}
and the $U$-dependent term
\begin{equation}
\begin{split}
    Q_U &= \sum_x \frac{\ri}{2} \Big\{  - \psi^\dagger_1(x) U(x) \psi_2(x+1) \\
    &\quad + \psi^\dagger_2(x+1) U^\dagger(x) \psi_1(x) \Big\} 
\end{split}
\end{equation}
for later use.
The axial charge commutes with the fermion Hamiltonian, $[H_f,Q_5] = 0$, but does not commute with the gauge Hamiltonian, $[H_g,Q_5] \neq 0$.
Thus it is conserved only for $e=0$.
The gauge interaction violates axial charge conservation, $\frac{d}{dt} Q_5 = \ri [H,Q_5] \neq 0$, and results in the axial anomaly \eqref{eqAC} in the U(1) limit.
In Ref.~\cite{Hidaka:2025ram}, it was analytically proven that, if the axial charge satisfies $[H_f,Q_5] = 0$, infinitesimal time evolution can reproduce the axial anomaly \eqref{eqAC}.
This property is useful for quantum computing because simulations can be implemented using very shallow circuits.
(Note that, for open boundary conditions, $[H_f,Q_5] = 0$ is violated and axial symmetry is artificially broken.
Periodic boundary conditions are essential for our simulation.)

We simulated the Hamiltonian time evolution of the axial charge \eqref{eqQ5}.
The right panel of Fig.~\ref{figschematic} shows an overview of our quantum circuit.
The circuit was simplified in various ways to avoid device limitations.
Each step is explained below and further details are explained in the Appendix.

\section{Initial state preparation}

The initial state is the non-interacting vacuum given by a direct product of gauge and fermion state vectors, 
\begin{equation}
    |\Psi(0)\rangle = |g\rangle \otimes |f\rangle .
\end{equation}
The gauge part corresponds to the weak coupling limit
\begin{equation}
    U(x)|g\rangle = |g\rangle \quad {\rm for} \ \forall x .
\end{equation}
The gauge register for each link is composed of $\log_2 N$ qubits, and all the qubits are initialized to $|0\rangle$ in the $U$-diagonal basis.
The fermion part is constructed by solving the non-interacting Wilson-Dirac equation and occupying the Dirac sea, i.e., all the single-particle states with negative energy.
(Zero-energy single-particle states are doubly degenerate and one of them must be occupied in the charge neutral vacuum.
The occupied state is selected by introducing an infinitesimal fermion mass and then taking the massless limit.)
The vacuum satisfies $Q|f\rangle = 0$ and $Q_5 |f\rangle=0$ and its total energy is 
$H_f |f\rangle = - \sum_k 2 \sin \left( \frac{\pi}{L}k \right)|f\rangle$.
The momentum space representation is
\begin{equation}
\label{eqfp}
    |f'\rangle = \prod_k \frac{1}{\sqrt{2}} \{\psi_1^\dagger(k) -e^{\ri\theta(k)}\psi_2^\dagger(k)\} |{\rm empty}\rangle ,
\end{equation}
with the phase angle
\begin{equation}
    \theta(k) = 
    \left\{
\begin{array}{ll}
0 & (k=0) \\
\frac{\pi}{2} -\frac{\pi}{L}k & (k= 1,2,\cdots,L-1)
\end{array}
\right.
\end{equation}
and the empty state $|{\rm empty}\rangle$.
The coordinate space representation is obtained by
\begin{equation}
\label{eqfx}
    |f\rangle = {\rm FFT}^\dagger_2 {\rm FFT}^\dagger_1 |f'\rangle
\end{equation}
where ${\rm FFT}^\dagger_i$ is the (inverse) fermionic Fourier transform for the $i$-th spinor component.
The fermionic Fourier transform is implemented using a circuit of sequential basis rotations~\cite{Wecker:2015fib}.

\section{Time evolution}

Starting from the above initial state, we simulated time evolution including the gauge interaction.
For $t>0$, the gauge field consists of an external classical background and quantum fluctuations.
The gauge link operator is replaced as $U(x) \to U(x) e^{\ri A_{\rm ext}}$ with a spatially uniform background $A_{\rm ext}$.
In ${\mathbb Z}_N$ gauge theory, the external field can take discrete values $A_{\rm ext}=\frac{2\pi}{N}, \frac{4\pi}{N}, \cdots, \frac{2\pi(N-1)}{N}$.
Such discrete jumps become continuous as $N\to \infty$, and lead to the standard setup $A_{\rm ext}=eE_{\rm ext}t$, which is used to derive the axial anomaly \eqref{eqAC}.

Time evolution is implemented by a single second-order Suzuki-Trotter step
\begin{equation}
    |\Psi(\delta t)\rangle = e^{-\ri H_f \frac{\delta t}{2}}e^{-\ri H_g \delta t}e^{-\ri H_f \frac{\delta t}{2}}|\Psi(0)\rangle 
\end{equation}
with a small time step $\delta t$.
In taking the expectation value of $Q_5$, the fermionic evolution operators $e^{-\ri H_f \frac{\delta t}{2}}$ can be omitted because $Q_5$ commutes with $H_f$ and $|\Psi(0)\rangle$ is an eigenstate of $H_f$,
\begin{equation}
   \langle \Psi(\delta t)| Q_5 |\Psi(\delta t)\rangle = \langle \Psi(0)| e^{\ri H_g \delta t} Q_5 e^{-\ri H_g \delta t}|\Psi(0)\rangle .
\end{equation}
In applying the evolution operator $e^{-\ri H_g \delta t}$, the gauge register is transformed from the $U$-diagonal basis to the $\Pi$-diagonal basis via the quantum Fourier transform (QFT) \cite{Nielsen:2012yss}.
After executing the evolution circuit, the basis is transformed back to the $U$-diagonal one by its conjugate ${\rm QFT}^\dagger$.

\section{Measurement}

Qubits are measured to compute the expectation value at the end of the simulation.
We can analytically show $\langle Q_5\rangle=\langle Q_M\rangle=\langle Q_U\rangle=0$ at $t=0$.
Since $\langle Q_M\rangle$ does not evolve in a single Suzuki-Trotter step due to $U$ independence, it is still zero at $t=\delta t$.
The axial charge at $t=\delta t$ is therefore given by
\begin{equation}
\begin{split}
    \langle Q_5 \rangle &= \langle Q_U \rangle \\
    &= \frac{\ri L}{2} \big\langle - \psi^\dagger_1(0) U(0)e^{\ri A_{\rm ext}} \psi_2(1) \\
    &\quad + \psi^\dagger_2(1) U^\dagger(0)e^{-\ri A_{\rm ext}} \psi_1(0) \big\rangle .
\end{split}
\end{equation}
The second equality is a consequence of translational invariance.
The number of gauge registers, as well as the number of gate operations, can be reduced by computing the expectation value only at $x=0$ and multiplying it by the volume factor $L$.

\section{Simulation results}

We utilized the trapped-ion quantum computer ``Reimei'', which was built by Quantinuum and deployed at RIKEN.
Reimei has 20 physical qubits and all-to-all connectivity.
We need to take four limits: the U(1) limit $N\to \infty$, the infinitesimal time limit $\delta t\to 0$, the infinite volume limit $L \to \infty$, and the continuum limit $e^2\to 0$.
The infinitesimal time and continuum limits are, however, equivalent because the circuits depend only on a combination $e^2\delta t$.
We executed 1000 shots for each of the 27 parameter sets $N=\{4,8,16\}$, $L=\{3,4,5\}$, and $e^2\delta t = \{0.1,0.2,0.3\}$.
The number of qubits scales as $N_{\rm bit}=2L+\log_2N$. 

\begin{figure}[htb]
    \centering
    \includegraphics[width=\linewidth]{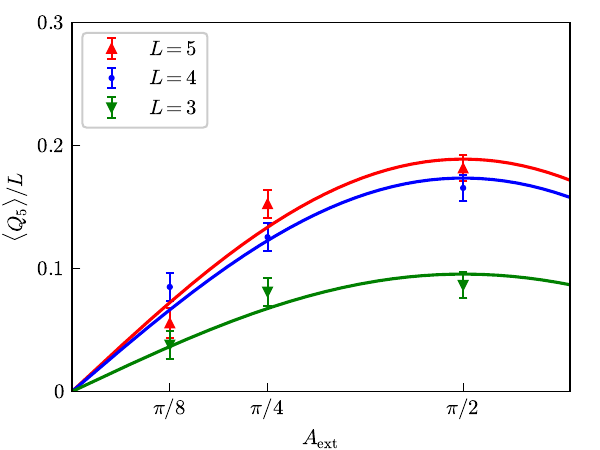}
    \caption{\label{figZN}
    Axial charge $\langle Q_5 \rangle$ produced by the external gauge field $A_{\rm ext}=\frac{2\pi}{N}$.
    The data of $e^2 \delta t = 0.1$ are shown.
    Error bars represent statistical uncertainties due to finite shots.}
\end{figure}

\begin{figure}[htb]
    \centering
    \includegraphics[width=\linewidth]{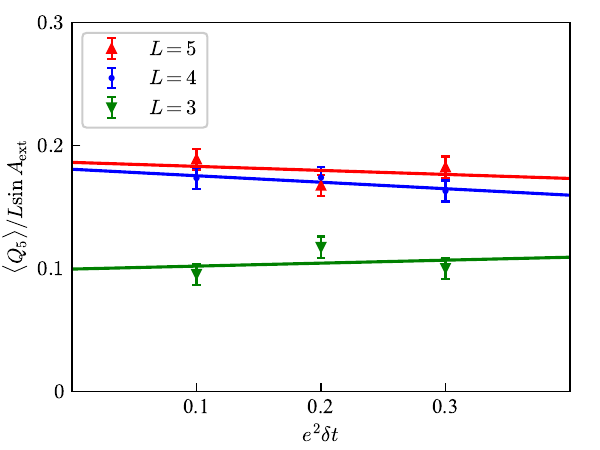}
    \caption{\label{figG}
    Infinitesimal time extrapolation $\delta t \to 0$.
    Error bars represent fitting uncertainties from Fig.~\ref{figZN}.}
\end{figure}

\begin{figure}[htb]
    \centering
    \includegraphics[width=\linewidth]{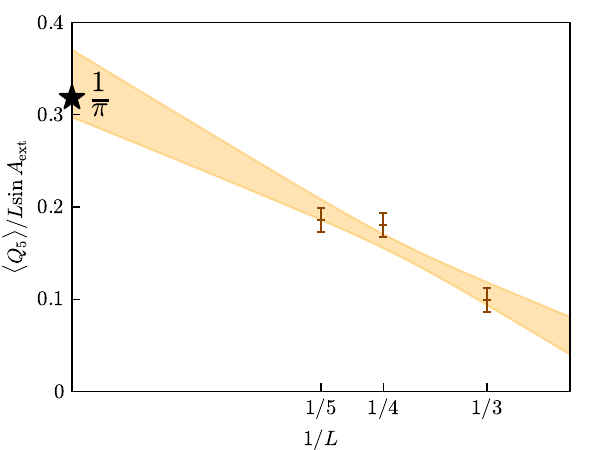}
    \caption{\label{figL}
    Infinite volume extrapolation $L \to \infty$.
    Error bars represent fitting uncertainties from Fig.~\ref{figG} and the error band represents the uncertainty of the linear extrapolation.
    The star denotes the correct anomaly coefficient $\frac{1}{\pi}$.}
\end{figure}

First, we estimated the U(1) limit from the simulations of ${\mathbb Z}_4$, ${\mathbb Z}_8$, and ${\mathbb Z}_{16}$ gauge theories.
The external field is set to a minimal nonzero value $A_{\rm ext} = \frac{2\pi}{N}$.
On a lattice, a simulation result must be a periodic function of $A_{\rm ext}$ because of the $2\pi$ periodicity of $e^{\ri A_{\rm ext}}$.
As shown in Fig.~\ref{figZN}, we used a fit function
\begin{equation}
\label{eqfitting}
    \langle Q_5 \rangle = C L \sin A_{\rm ext}
\end{equation}
and obtained best-fit values of the numerical coefficient $C$.
This estimates the anomaly coefficient of
\begin{equation}
    \langle Q_5 \rangle = C L A_{\rm ext} = C L eE_{\rm ext} \delta t
\end{equation}
in U(1) gauge theory.
Next, we extrapolated the obtained values to the infinitesimal time limit.
Given the weak dependence on $\delta t$ in Fig.~\ref{figG}, a linear extrapolation is justified.
Using the results of the infinitesimal time extrapolation, we finally took the infinite volume limit.
As shown in Fig.~\ref{figL}, the linear extrapolation yields the correct anomaly coefficient $\frac{1}{\pi}$ within the error band.
The final result after all the extrapolations is $C=0.33\pm0.04$.
This demonstrates the validity of our simulation.


No error mitigation was used in the above analysis.
This is due to the high fidelity of the device and the simplicity of the circuit.
The dominant source of noise is two-qubit gate errors, whose error rate is $p = 1.30 \times 10^{-3}$.
Owing to substantial simplification of the circuits, the number of two-qubit gates is $N_{\rm 2Q}=27$ for the minimal case (${\mathbb Z}_4$ with $L=3$) and $N_{\rm 2Q}=106$ for the maximal case (${\mathbb Z}_{16}$ with $L=5$).
The infidelity is therefore estimated by assuming a depolarizing channel as $1-(1-p)^{N_{\rm 2Q}}=3.5\%\text{--}12.9\%$, which is smaller than the fitting uncertainty arising from statistical errors.

\section{Extensibility and scalability}

We have demonstrated that the axial anomaly can be simulated on a current quantum computer without error mitigation.
The required computational resources are surprisingly small.
The maximum number of qubits used in our simulations is $N_{\rm bit}=14$.
Such small-scale simulations are tractable on classical computers, and we checked the consistency of the results there.
The same analysis can be extended to large-scale simulations involving $N_{\rm bit}>100$, which are beyond the reach of classical computers, although error mitigation will be necessary.

The analysis can be extended in three directions.
i) As the volume $L$ increases, the number of qubits increases linearly.
The two-qubit gate count of the fast fermionic Fourier transform scales as $O(L\log L)$, while that of a Pauli string scales as $O(L)$ in the Jordan-Wigner encoding and $O(\log L)$ in the Bravyi-Kitaev encoding.
Naive multiplication gives $O(L^2\log L)$ and $O(L\log^2 L)$, respectively, although this often overestimates the actual cost.
ii) As the gauge field dimension $N$ increases, the number of qubits increases logarithmically.
The two-qubit gate count is $O(\log N)$ for QFT and remains subleading for the gauge Hamiltonian evolution.
iii) Although the infinitesimal time and continuum limits are equivalent for a single Suzuki-Trotter step, they are independent in general.
The systematic error can be estimated by employing higher-order formulas or multiple steps.
This broadens the causal cone in time evolution and increases the computational cost in a nontrivial manner.

\begin{acknowledgments}
We acknowledge support from the RIKEN TRIP initiative (RIKEN Quantum).
T.~H.~was supported by JSPS KAKENHI Grant No.~JP24K00630 and JP25K01002.
A.~Y.~was supported by JSPS KAKENHI Grant No.~JP25K07295.
This work was based on results obtained from a project, JPNP20017, commissioned by the New Energy and Industrial Technology Development Organization (NEDO).
\end{acknowledgments}

\onecolumngrid
\appendix*

\section{Simulation details}

In this Appendix, we describe the details of the quantum circuits used in our simulations.
Circuit implementations are defined up to a physically irrelevant global phase.

\subsection{Fermionic vacuum in momentum space}

We adopt the Jordan-Wigner transformation
\begin{equation}
    \psi^\dagger_i(x) = \frac{X_i(x)+\ri Y_i(x)}{2} \prod_{(j,y)<(i,x)} \left\{ -Z_j(y) \right\} ,
\end{equation}
where the ordering is taken along the one-dimensional chain of fermions.
In this convention, $|0\rangle$ is an occupied state and $|1\rangle$ is an empty state.
We define $C(k)$ as the operator to create a non-interacting Wilson fermion with momentum $k$ and energy $- 2 \sin \left( \frac{\pi}{L}k \right)$.
It acts on the initial fermion qubits $|f_2(k)f_1(k)\rangle=|00\rangle$ as
\begin{equation}
    C(k)|00\rangle = \frac{1}{\sqrt{2}} \{\psi_1^\dagger(k) -e^{\ri\theta(k)}\psi_2^\dagger(k)\} |11\rangle
\end{equation}
with
\begin{equation}
    e^{\ri\theta(k)} = 
    \left\{
\begin{array}{ll}
1 & (k=0) \\
\frac{1-e^{-\ri\frac{2\pi k}{L}}}{|1-e^{-\ri\frac{2\pi k}{L}}|} & (k= 1,2,\cdots,L-1).
\end{array}
\right.
\end{equation}
The circuit is given by
\begin{equation}
    C(k) =
\begin{quantikz}[column sep=1.0em, row sep=0.4em] 
\lstick{$f_2(k)$} & \qw & \qw & \qw & \targ{} & \gate{X} & \qw & \qw \\
\lstick{$f_1(k)$} & \gate{X} & \gate{H} & \gate{R_Z(\theta)} & \ctrl{-1} & \qw & \qw & \qw
\end{quantikz}
.
\end{equation}

\subsection{Fermionic Fourier transform}

The fermionic Fourier transform is defined by
\begin{equation}
    {\rm FFT}^\dagger_i=\exp\left\{ -\sum_{xk} \psi_i^\dagger(x)\left(\ln F \right)_{xk} \psi_i(k) \right\} ,
\end{equation}
where $F$ is the Fourier matrix
\begin{equation}
    \psi_i(x) = \sum_k F_{xk} \psi_i(k) = \frac{1}{\sqrt{L}}\sum_k e^{\ri \frac{2\pi kx}{L}}\psi_i(k)  .
\end{equation}
The Fourier matrix $F$ can be decomposed into $L(L-1)/2$ Givens rotations and a diagonal phase giving $\det F \neq 1$.

For $L=3$, the matrix is decomposed into three rotations and a diagonal phase
\begin{equation}
\label{eqGivens3}
F=
\frac{1}{\sqrt{3}}
\begin{pmatrix}
1 & 1 & 1 \\
1 & \omega & \omega^2 \\
1 & \omega^2 & \omega
\end{pmatrix}
= 
\begin{pmatrix}
\frac{1}{\sqrt{2}} & -\frac{1}{\sqrt{2}} & 0 \\
\frac{1}{\sqrt{2}} & \frac{1}{\sqrt{2}} & 0 \\
0 & 0 & 1
\end{pmatrix}
\begin{pmatrix}
\frac{\sqrt{2}}{\sqrt{3}} & 0 & -\frac{1}{\sqrt{3}} \\
0 & 1 & 0 \\
\frac{1}{\sqrt{3}} & 0 & \frac{\sqrt{2}}{\sqrt{3}}
\end{pmatrix}
\begin{pmatrix}
1 & 0 & 0 \\
0 & -\frac{\ri}{\sqrt{2}}\omega^2 & -\frac{1}{\sqrt{2}}\omega \\
0 & \frac{1}{\sqrt{2}}\omega^2 & \frac{\ri}{\sqrt{2}}\omega
\end{pmatrix}
\begin{pmatrix}
1 & 0 & 0 \\
0 & 1 & 0 \\
0 & 0 & -\ri
\end{pmatrix}
\end{equation}
with $\omega=\exp(\ri \frac{2\pi}{L})$.
For convenience, we introduce some notation: the real rotation matrix in the $i,j$ plane, $r_{ij}(\theta)$, the complex rotation matrix in the $i,j$ plane, $u_{ij}$, and the diagonal matrix $d(\alpha,\beta,\gamma)={\rm diag}(e^{\ri \alpha},e^{\ri \beta},e^{\ri \gamma})$.
Using this notation, Eq.~\eqref{eqGivens3} is rewritten as
\begin{equation}
F= r_{01} (\tfrac{\pi}{4}) r_{02} (\arccos\tfrac{\sqrt{2}}{\sqrt{3}}) u_{12} d(0,0,-\tfrac{\pi}{2}) .
\end{equation}
Since an SU(2) matrix is generally written as
\begin{equation}
\begin{pmatrix}
 ce^{\ri\alpha} & - se^{\ri\beta} \\
 se^{-\ri\beta} & ce^{-\ri\alpha}
\end{pmatrix}
= 
\begin{pmatrix}
 1 & 0 \\
 0 & e^{-\ri(\alpha+\beta)}
\end{pmatrix}
\begin{pmatrix}
 c & - s \\
 s & c
\end{pmatrix}
\begin{pmatrix}
 e^{\ri\alpha} & 0 \\
 0 & e^{\ri\beta}
\end{pmatrix},
\end{equation}
the complex rotation is further decomposed as
\begin{equation}
u_{12} = d(0,0,\tfrac{\pi}{2}) r_{12} (\tfrac{\pi}{4}) d(0,\tfrac{5\pi}{6},\tfrac{2\pi}{3}).
\end{equation}
Thus the final form of the decomposition is
\begin{equation}
F = r_{01} (\tfrac{\pi}{4}) r_{02} (\arccos\tfrac{\sqrt{2}}{\sqrt{3}}) d(0,0,\tfrac{\pi}{2}) r_{12} (\tfrac{\pi}{4}) d(0,\tfrac{5\pi}{6},\tfrac{2\pi}{3}-\tfrac{\pi}{2}) .
\end{equation}
The fermionic Fourier transform is constructed by using these rotation angles and diagonal phases;
\begin{equation}
    {\rm FFT}^\dagger_1=R_{01} (\tfrac{\pi}{4}) R_{02} (\arccos\tfrac{\sqrt{2}}{\sqrt{3}}) D(0,0,\tfrac{\pi}{2}) R_{12} (\tfrac{\pi}{4}) D(0,\tfrac{5\pi}{6},\tfrac{2\pi}{3}-\tfrac{\pi}{2})
\end{equation}
with 
\begin{align}
    R_{01} (\theta_{01}) &= \exp\left[-\theta_{01}\{\psi^\dagger_1(0)\psi_1(1)-\psi^\dagger_1(1)\psi_1(0)\}\right] \\
    R_{02} (\theta_{02}) &= \exp\left[-\theta_{02}\{\psi^\dagger_1(0)\psi_1(2)-\psi^\dagger_1(2)\psi_1(0)\}\right] \\
    R_{12} (\theta_{12}) &= \exp\left[-\theta_{12}\{\psi^\dagger_1(1)\psi_1(2)-\psi^\dagger_1(2)\psi_1(1)\}\right] \\
    D(\phi_0,\phi_1,\phi_2) &= \exp\left[\ri\phi_0 \psi^\dagger_1(0)\psi_1(0)+\ri\phi_1 \psi^\dagger_1(1)\psi_1(1)+\ri\phi_2 \psi^\dagger_1(2)\psi_1(2) \right] .
\end{align}
The same goes for ${\rm FFT}^\dagger_2$ except that $\psi_1^\dagger$ and $\psi_1$ are replaced by $\psi_2^\dagger$ and $\psi_2$, respectively.
By the Jordan-Wigner transformation, we get
\begin{align}
    R_{01} (\theta_{01}) &= \exp\left[-\ri\frac{\theta_{01}}{2}\{X_1(0)Y_1(1)-X_1(1)Y_1(0)\}Z_2(1)\right] \\
    \begin{split}\label{eqR02}
        R_{02} (\theta_{02}) &= \exp\left[-\ri\frac{\theta_{02}}{2}\{X_1(0)Y_1(2)-X_1(2)Y_1(0)\}Z_2(1)Z_1(1)Z_2(2)\right] \\
        &= \exp\left[-\ri\frac{\theta_{02}}{2}\{X_1(0)Y_1(2)-X_1(2)Y_1(0)\}Z_2(0)\right]
    \end{split}\\
    R_{12} (\theta_{12}) &= \exp\left[-\ri\frac{\theta_{12}}{2}\{X_1(1)Y_1(2)-X_1(2)Y_1(1)\}Z_2(2)\right] \\
    D(\phi_0,\phi_1,\phi_2) &= \exp\left[\ri \frac{\phi_0}{2} \{Z_1(0)+1\} + \ri \frac{\phi_1}{2} \{Z_1(1)+1\} + \ri \frac{\phi_2}{2} \{Z_1(2)+1\} \right]
\end{align}
In Eq.~\eqref{eqR02}, we used the equality $Z_2(1)Z_1(1)Z_2(2)=(-1)^{Q+L-1} Z_2(0)$, which is derived from the periodicity of the lattice.
The Givens rotation operators $R_{ij}$ are implemented by the circuit
\begin{equation}
e^{-\ri\frac{\theta}{2}(X_0Y_2-Y_0X_2)Z_1} =
\begin{quantikz}[column sep=1.0em, row sep=0.4em] 
\lstick{$f_0$} & \gate{S} & \targ{} & \qw & \gate{H} & \targ{} & \gate{H} & \gate{S} & \qw & \targ{} & \qw & \gate{Z} & \qw & \qw \\
\lstick{$f_1$} & \gate{H} & \ctrl{-1} & \ctrl{1} & \gate{R_X(\theta)} & \ctrl{-1} & \gate{Z} & \ctrl{1} & \gate{R_X(\theta)} & \ctrl{-1} & \ctrl{1} & \gate{H} & \qw & \qw \\
\lstick{$f_2$} & \qw & \qw & \targ{} & \gate{H} & \qw & \qw & \targ{} & \gate{H} & \gate{S} & \targ{} & \gate{S^\dagger} & \qw & \qw
\end{quantikz}
\end{equation}
and the diagonal phase operator $D$ is implemented by $R_Z$ gates on individual qubits.

For $L=4$, the Fourier matrix is decomposed as
\begin{equation}
\frac{1}{2}
\begin{pmatrix}
1 & 1 & 1 & 1\\
1 & \ri & -1 & -\ri \\
1 & -1 & 1 & -1 \\
1 & -\ri & -1 & \ri
\end{pmatrix}
= r_{01} (\tfrac{\pi}{4}) r_{02} (\arccos\tfrac{\sqrt{2}}{\sqrt{3}}) r_{03} (\tfrac{\pi}{6}) r_{12} (\tfrac{\pi}{3}) r_{23} (-\arccos\tfrac{\sqrt{2}}{\sqrt{3}}) u_{13} d(0,0,0,-\tfrac{\pi}{2})
\end{equation}
with
\begin{equation}
u_{13}=
\begin{pmatrix} 
1 & 0 & 0 & 0 \\ 
0 & -\frac{1}{\sqrt{2}} & 0 & -\frac{\ri}{\sqrt{2}} \\ 
0 & 0 & 1 & 0 \\ 
0 & -\frac{\ri}{\sqrt{2}} & 0 & -\frac{1}{\sqrt{2}} 
\end{pmatrix}.
\end{equation}
Based on this decomposition, we can construct the operators and the circuit as in the case above, although we omit their explicit forms.
A difference is that one of the Givens rotation operators contains a string of $ZZZ$ as
\begin{equation}
    R_{02} (\theta_{02}) = \exp\left[-\ri\frac{\theta_{02}}{2}\{X_1(0)Y_1(2)-X_1(2)Y_1(0)\}Z_2(1)Z_1(1)Z_2(2)\right] .
\end{equation}
The $ZZZ$ string can be implemented by sandwiching by CNOT ladders,
\begin{equation}
e^{-\ri\frac{\theta}{2}(X_0Y_4-Y_0X_4)Z_1Z_2Z_3} =
\begin{quantikz}[column sep=1.0em, row sep={2em,between origins}]
\lstick{$f_0$} & \qw & \qw & \gate[wires=3]{e^{-\ri\frac{\theta}{2}(X_0Y_4-Y_0X_4)Z_1}} & \qw & \qw & \qw \\ 
\lstick{$f_1$} & \targ{} & \targ{} & \qw & \targ{} & \targ{} & \qw \\
\lstick{$f_4$} & \qw & \qw & \qw & \qw & \qw & \qw \\
\lstick{$f_2$} & \qw & \ctrl{-2} & \qw  & \ctrl{-2} & \qw & \qw \\
\lstick{$f_3$} & \ctrl{-3} & \qw & \qw  & \qw & \ctrl{-3} & \qw
\end{quantikz}
\end{equation}
Another rotation operator contains a string of $ZZZZZ$ but it can be reduced as
\begin{equation}
\begin{split}
    R_{03} (\theta_{03}) &= \exp\left[-\ri\frac{\theta_{03}}{2}\{X_1(0)Y_1(3)-X_1(3)Y_1(0)\}Z_2(1)Z_1(1)Z_2(2)Z_2(1)Z_2(3)\right] \\
    &= \exp\left[\ri\frac{\theta_{03}}{2}\{X_1(0)Y_1(3)-X_1(3)Y_1(0)\}Z_2(0)\right]
\end{split}
\end{equation}
because of $Z_2(1)Z_1(1)Z_2(2)Z_1(2)Z_2(3)=(-1)^{Q+L-1} Z_2(0)$.

For $L=5$, the decomposition is
\begin{align}
& \frac{1}{\sqrt{5}}
\begin{pmatrix}
1 & 1 & 1 & 1 & 1\\
1 & \omega & \omega^2 & \omega^3 & \omega^4 \\
1 & \omega^2 & \omega^4 & \omega & \omega^3 \\
1 & \omega^3 & \omega & \omega^4 & \omega^2 \\
1 & \omega^4 & \omega^3 & \omega^2 & \omega
\end{pmatrix}
= r_{01} (\tfrac{\pi}{4}) r_{02} (\arccos\tfrac{\sqrt{2}}{\sqrt{3}}) r_{03} (\tfrac{\pi}{6}) r_{04} (\arccos\tfrac{2}{\sqrt{5}}) u_{12} u_{13} u_{14} u_{23} u_{24} u_{34} d(0,0,0,0,\pi)
\end{align}
with
\begin{align}
u_{12} &=
\begin{pmatrix}
1 & 0 & 0 & 0 & 0 \\
0 & \frac{\sqrt{3}}{\sqrt{15-\sqrt{5}}}(\omega-1) & -\frac{1}{\sqrt{15-\sqrt{5}}} (3\omega^3+\omega^2+\omega) & 0 & 0 \\
0 & \frac{1}{\sqrt{15-\sqrt{5}}} (3\omega^2+\omega^3+\omega^4) & \frac{\sqrt{3}}{\sqrt{15-\sqrt{5}}}(\omega^4-1) & 0 & 0 \\
0 & 0 & 0 & 1 & 0 \\
0 & 0 & 0 & 0 & 1
\end{pmatrix}
\\
u_{13} &=
\begin{pmatrix}
1 & 0 & 0 & 0 & 0 \\
0 & \frac{\sqrt{2(15 - \sqrt{5})}}{3\sqrt{5}} & 0 & - \frac{1}{3\sqrt{5}}\left( 4\omega^{2} + \omega \right) & 0 \\
0 & 0 & 1 & 0 & 0 \\
0 & \frac{1}{3\sqrt{5}}\left( 4\omega^{3} + \omega^4 \right) & 0 & \frac{\sqrt{2(15 - \sqrt{5})}}{3\sqrt{5}} & 0 \\
0 & 0 & 0 & 0 & 1
\end{pmatrix}
\\
u_{14} &=
\begin{pmatrix}
1 & 0 & 0 & 0 & 0 \\
0 & \frac{\sqrt{3}}{2} & 0 & 0 & -\frac{1}{2}\omega\\
0 & 0 & 1 & 0 & 0\\
0 & 0 & 0 & 1 & 0\\
0 & \frac{1}{2}\omega^4 & 0 & 0 & \frac{\sqrt{3}}{2}
\end{pmatrix}
\\
u_{23} &=
\begin{pmatrix}
1 & 0 & 0 & 0 & 0 \\
0 & 1 & 0 & 0 & 0 \\
0 & 0 & \frac{\sqrt{3}}{\sqrt{15-\sqrt{5}}}(\omega-1) & -\frac{1}{\sqrt{15-\sqrt{5}}}(3\omega^{4}+\omega^{3}+\omega^{2}) & 0 \\
0 & 0 & \frac{1}{\sqrt{15-\sqrt{5}}}(3\omega+\omega^2+\omega^3) & \frac{\sqrt{3}}{\sqrt{15-\sqrt{5}}}(\omega^4-1) & 0 \\
0 & 0 & 0 & 0 & 1
\end{pmatrix}
\\
u_{24} &=
\begin{pmatrix}
1 & 0 & 0 & 0 & 0 \\
0 & 1 & 0 & 0 & 0 \\
0 & 0 & \frac{\sqrt{2}}{\sqrt{3}} & 0 & -\frac{1}{\sqrt{3}}\omega^2  \\
0 & 0 & 0 & 1 & 0\\
0 & 0 & \frac{1}{\sqrt{3}}\omega^3 & 0  & \frac{\sqrt{2}}{\sqrt{3}}
\end{pmatrix}
\\
u_{34} &=
\begin{pmatrix}
1 & 0 & 0 & 0 & 0 \\
0 & 1 & 0 & 0 & 0 \\
0 & 0 & 1 & 0 & 0 \\
0 & 0 & 0 & \frac{\sqrt{10}}{15-\sqrt{5}}(3\omega+3\omega^2+2\omega^4) & -\frac{\sqrt{2}}{15-\sqrt{5}}(1+8\omega^3+\omega) \\
0 & 0 & 0 & \frac{\sqrt{2}}{15-\sqrt{5}}(1+8\omega^2+\omega^4) & \frac{\sqrt{10}}{15-\sqrt{5}}(3\omega^4+3\omega^3+2\omega) 
\end{pmatrix} .
\end{align}
The corresponding operators are written by up to a string of $ZZZ$, so practical implementation is the same as $L=4$. 

\subsection{Quantum Fourier transform}

Let $|g_0g_1\cdots\rangle$ denote a gauge register: $|g_0g_1\rangle$ for ${\mathbb Z}_4$, $|g_0g_1g_2\rangle$ for ${\mathbb Z}_8$, and $|g_0g_1g_2g_3\rangle$ for ${\mathbb Z}_{16}$.
The gauge register is initially prepared in the $U$-diagonal basis as
\begin{equation}
    U|00\cdots\rangle = \exp\left\{ \ri 2\pi \left( \frac12 g_0 + \frac14 g_1 + \cdots \right) \right\}|00\cdots\rangle = |00\cdots\rangle
\end{equation}
Before the time evolution, it is transformed to the $\Pi$-diagonal basis by the QFT.
Since all the qubits are initially set to $|0\rangle$, we can easily get
\begin{equation}
    {\rm QFT} |00\cdots\rangle = \left\{ \frac{1}{\sqrt{2}} (|0\rangle+|1\rangle) \right\}^{\log_2 N} = H_0H_1\cdots |00\cdots\rangle .
\end{equation}
The circuit is just parallel Hadamard gates.
After the time evolution, it is transformed to the $U$-diagonal basis by ${\rm QFT}^\dagger$ in a general form.
Although the circuits are well-known, we write them down for completeness:
\begin{equation}
    {\rm QFT}^\dagger = 
\begin{quantikz}[column sep=1.0em, row sep=0.4em] 
\lstick{$g_0$} & \qw & \gate{R_Z(-\tfrac{\pi}{2})} & \gate{H} & \qw & \qw \\
\lstick{$g_1$} & \gate{H} & \ctrl{-1} & \qw & \qw & \qw
\end{quantikz}
\end{equation}
 for ${\mathbb Z}_4$,
\begin{equation}
    {\rm QFT}^\dagger =
\begin{quantikz} [column sep=1.0em, row sep=0.4em] 
\lstick{$g_0$} & \qw & \qw & \qw & \gate{R_Z(-\tfrac{\pi}{4})} & \gate{R_Z(-\tfrac{\pi}{2})} & \gate{H} & \qw & \qw \\ 
\lstick{$g_1$} & \qw & \gate{R_Z(-\tfrac{\pi}{2})} & \gate{H} & \qw & \ctrl{-1} & \qw & \qw & \qw \\ 
\lstick{$g_2$} & \gate{H} & \ctrl{-1} & \qw & \ctrl{-2} & \qw & \qw & \qw & \qw
\end{quantikz}
\end{equation}
for ${\mathbb Z}_8$, and
\begin{equation}
    {\rm QFT}^\dagger = 
\begin{quantikz}[column sep=0.5em, row sep=0.4em] 
\lstick{$g_0$} & \qw & \qw & \qw & \qw & \qw & \qw &
\gate{R_Z(-\tfrac{\pi}{8})} &
\gate{R_Z(-\tfrac{\pi}{4})} &
\gate{R_Z(-\tfrac{\pi}{2})} &
\gate{H} & \qw & \qw \\
\lstick{$g_1$} & \qw & \qw & \qw &
\gate{R_Z(-\tfrac{\pi}{4})} &
\gate{R_Z(-\tfrac{\pi}{2})} &
\gate{H} & \qw & \qw & 
\ctrl{-1} & \qw & \qw & \qw \\
\lstick{$g_2$} & \qw &
\gate{R_Z(-\tfrac{\pi}{2})} &
\gate{H} & \qw &
\ctrl{-1} & \qw & \qw &
\ctrl{-2} & \qw & \qw & \qw & \qw \\
\lstick{$g_3$} &
\gate{H} &
\ctrl{-1} & \qw &
\ctrl{-2} & \qw & \qw &
\ctrl{-3} & \qw & \qw & \qw & \qw & \qw
\end{quantikz}
\end{equation}
for ${\mathbb Z}_{16}$.
Note that the qubit order is swapped after ${\rm QFT}^\dagger$.

\subsection{Time evolution}

The time evolution operator is
\begin{equation}
e^{-\ri H_g \delta t} = \exp\left[ \ri \sum_x \frac{e^2\delta t}{2} \{ \Pi(x)+\Pi^\dagger(x) \} \right] .
\end{equation}
We only consider the time evolution at $x=0$ and omit the argument $x$.
In the $\Pi$-diagonal basis, 
\begin{equation}
\begin{split}
\frac{1}{2} ( \Pi+\Pi^\dagger ) 
&= \cos \left\{ 2\pi \left( \frac12 g_0 + \frac14 g_1 + \cdots \right) \right\}
\\
&= \cos \left\{ \pi \left( \frac12 (1-Z_0) + \frac14 (1-Z_1) +\cdots \right) \right\}
\end{split}
\end{equation}
is a function of Pauli-$Z$ operators.
This can be expanded in a finite power series of Pauli-$Z$ operators.
In ${\mathbb Z}_4$ gauge theory, the expansion is
\begin{equation}
    \frac{1}{2} ( \Pi+\Pi^\dagger ) = \frac12 Z_0 ( 1 + Z_1 )
\end{equation}
and the circuit is simply given by
\begin{equation}
e^{\ri \frac{\theta}{2} Z_0(1+Z_1)} =
\begin{quantikz}[column sep=1.0em, row sep=0.4em] 
\lstick{$g_0$} & \qw & \gate{R_Z(-\theta)} & \ctrl{1} & \qw \\ 
\lstick{$g_1$} & \qw & \gate{R_Z(-\theta)} & \targ{}  & \qw \ .
\end{quantikz}
\end{equation}
The expansion for ${\mathbb Z}_8$ is
\begin{equation}
    \frac{1}{2} ( \Pi+\Pi^\dagger ) = \frac14 Z_0 \left\{ 1 + (1+\sqrt{2})Z_1 + Z_2 + (1-\sqrt{2})Z_1Z_2 \right\}
\end{equation}
and the expansion for ${\mathbb Z}_{16}$ is
\begin{equation}
\begin{split}
    \frac{1}{2} ( \Pi+\Pi^\dagger ) = \frac18 Z_0 \Big\{ 1 &+ (1+c_++c_-+\sqrt{2})Z_1 + (1+c_+-c_-)Z_2 + Z_3 + (1-\sqrt{2})Z_1Z_2 \\
    &+ (1-c_+-c_-+\sqrt{2})Z_1Z_3 + (1-c_++c_-)Z_2Z_3 +  (1-\sqrt{2})Z_1Z_2Z_3 \Big\}
\end{split}
\end{equation}
with $c_+=2\cos(\tfrac{\pi}{8})=\sqrt{2+\sqrt{2}}$ and $c_-=2\cos(\tfrac{3\pi}{8})=\sqrt{2-\sqrt{2}}$.

\subsection{Measurement basis transformation}

To compute the expectation value of the operator
\begin{equation}
    q = \frac{\ri}{2} \left\{  - \psi^\dagger_1(0) U(0)e^{\ri A_{\rm ext}} \psi_2(1) + \psi^\dagger_2(1) U^\dagger(0)e^{-\ri A_{\rm ext}} \psi_1(0) \right\} ,
\end{equation}
we finally perform a basis transformation to the measurement basis.
For example, the circuit to diagonalize $q$ for ${\mathbb Z}_{16}$ is
\begin{equation}
B =
\begin{quantikz}[column sep=0.5cm, row sep=0.4cm]
\lstick{$f_1$} & \qw & \qw & \qw & \qw & \qw & \qw & \ctrl{5} & \gate{H} & \ctrl{5} & \qw \\
\lstick{$g_0$} & \qw & \qw & \qw & \qw & \qw & \ctrl{4} & \qw & \qw & \qw & \qw \\
\lstick{$g_1$} & \qw & \qw & \qw & \qw & \ctrl{3} & \qw & \qw & \qw & \qw & \qw \\
\lstick{$g_2$} & \qw & \qw & \qw & \ctrl{2} & \qw & \qw & \qw & \qw & \qw & \qw \\
\lstick{$g_3$} & \qw & \qw & \ctrl{1} & \qw & \qw & \qw & \qw & \qw & \qw & \qw \\
\lstick{$f_2$} & \gate{S^\dagger} & \gate{R_Z(A_{\mathrm{ext}})} & \gate{R_Z(\pi)}
               & \gate{R_Z(\frac{\pi}{2})} & \gate{R_Z(\frac{\pi}{4})} & \gate{R_Z(\frac{\pi}{8})}
               & \targ{} & \qw & \targ{} & \qw ,
\end{quantikz}
\end{equation}
where the qubit order in the gauge register is swapped after ${\rm QFT}^\dagger$.
Since the circuit absorbs the U(1) phase of the gauge register into $f_2$, only the fermion qubits $|f_1 f_2\rangle$ are measured.

\subsection{Computational cost}

Table \ref{tabcost} summarizes the computational cost to simulate the circuits.
The circuits were compiled by the TKET compiler with $\texttt{optimisation\_level}=3$~\cite{Sivarajah2020}.

\setlength{\tabcolsep}{8pt}
\begin{table}[!ht]
    \centering
    \begin{tabular}{cc|ccc|}
         & & $N_{\rm bit}$ & $N_{\rm 2Q}$ & $D_{\rm 2Q}$ \\
         \hline
               & ${\mathbb Z}_4$    & 8  & 27 & 13 \\
         $L=3$ & ${\mathbb Z}_8$    & 9  & 33 & 13 \\
               & ${\mathbb Z}_{16}$ & 10 & 44 & 16 \\
         \hline
               & ${\mathbb Z}_4$    & 10 & 49 & 26 \\
         $L=4$ & ${\mathbb Z}_8$    & 11 & 55 & 27 \\
               & ${\mathbb Z}_{16}$ & 12 & 67 & 27 \\
         \hline
               & ${\mathbb Z}_4$    & 12 & 88 & 47 \\
         $L=5$ & ${\mathbb Z}_8$    & 13 & 94 & 46 \\
               & ${\mathbb Z}_{16}$ & 14 & 106 & 49 \\
         \hline
    \end{tabular}
    \caption{\label{tabcost}
    Circuit properties.
    The number of qubits is $N_{\rm bit}=2L({\rm fermion})+\log_2N ({\rm gauge})$. 
    The number $N_{\rm 2Q}$ and depth $D_{\rm 2Q}$ of ZZPhase gates on the compiled circuits are also listed.
    The depth is defined by the number of gates in the longest acyclic path through the circuit from the qubit initializations to the final measurements.
    }
\end{table}

\subsection{Classical simulation}

\begin{figure}[htb]
    \centering
    \includegraphics[width=0.5\linewidth]{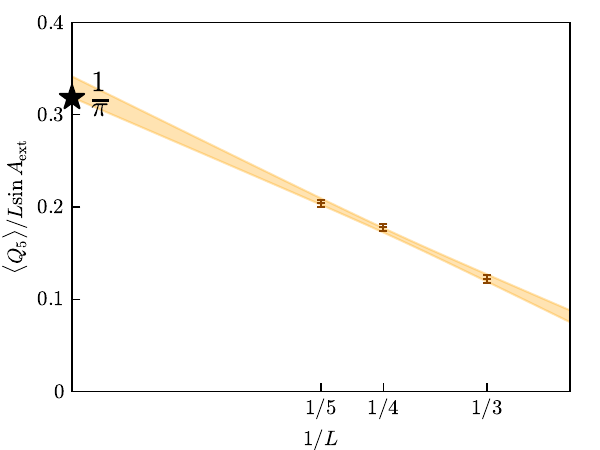}
    \caption{\label{figclassical}
    Classical simulation results corresponding to Fig.~4.
    The data are obtained by noiseless simulations with 10000 shots.}
\end{figure}

Since the computational cost is sufficiently small for the system sizes considered in this work, we cross-checked the validity of our analysis using a classical computer.
We executed the same circuits and performed the same fitting analysis as in the quantum simulations.
The classical simulations are free from device errors and have ten times more shots than the quantum simulations.
Figure \ref{figclassical} shows the final result of the infinite-volume extrapolation.
The classical simulations reproduce the anomaly coefficient $\frac{1}{\pi}$ with better accuracy, and thus support the analysis in the main text. 

\bibliography{paper}

\end{document}